# Strong gravitational field light deflection in binary systems containing a collapsed star

S. Campana[1,2]*, A. Parodi[3] and L. Stella[1,2]†
[1] *Osservatorio Astronomico di Brera, Via Bianchi 46, I-22055 Merate (LC), Italy*
[2] *Affiliated to I.C.R.A.*
[3] *Dipartimento di Fisica, Università di Roma "La Sapienza", P.le Aldo Moro, I-00185 Roma, Italy*

**ABSTRACT**
Large light deflection angles are produced in the strong gravitational field regions around neutron stars and black holes. In the case of binary systems, part of the photons emitted from the companion star towards the collapsed object are expected to be deflected in the direction of the earth. Based on a semi-classical approach we calculate the characteristic time delays and frequency shifts of these photons as a function of the binary orbital phase. The intensity of the strongly deflected light rays is reduced by many orders of magnitude, therefore making the observations of this phenomenon extremely difficult. Relativistic binary systems containing a radio pulsar and a collapsed object are the best available candidates for the detection of the strongly deflected photons. Based on the accurate knowledge of their orbital parameters, these systems allow to predict accurately the delays of the pulses along the highly deflected path, such that the sensitivity to very weak signals can be substantially improved through coherent summation over long time intervals. We discuss in detail the cases of PSR 1913+16 and PSR 1534+12 and find that the system geometry is far more promising for the latter. The observation of the highly deflected photons can provide a test of general relativity in an unprecedented strong field regime as well as a tight constraint on the radius of the collapsed object.

# 1 INTRODUCTION

Pulse timing measurements of the binary radio pulsar PSR 1913+16 over the last two decades (Hulse & Taylor 1975) have provided very accurate tests of crucial predictions of general relativity. These include the orbital period decay induced by the emission of gravitational radiation, the advance of periastron and time-dilation/gravitational redshift effects (e.g. Damour & Taylor 1992; Taylor 1992). As a consequence of these findings, alternative theories of gravity have either been ruled out, or severely constrained (e.g. Taylor et al. 1992). In the last few years several new binary systems containing a radio pulsar and a neutron star have been discovered: PSR 2127+11C in the globular cluster M15 (Anderson et al. 1990), PSR 1534+12 (Wolszczan 1991) and PSR 2303+46 (Thorsett et al. 1993). These systems are expected to provide even more stringent tests of gravitation theories in the near future. Despite these very important successes, all measured relativistic quantities involve gravitational fields with $2\,G\,M/R_{\rm min}\,c^2 \lesssim 10^{-5}$, where $R_{\rm min}$ is the minimum distance between the companion neutron star (of mass $M$) and the observed radio beam.

Here we suggest a possible way of testing a strong field prediction of general relativity in binary systems containing a neutron star or a black hole, namely light deflection by large angles ($\gtrsim 1$ rad) in the vicinity of the collapsed star ($R_{\rm min} \sim 3\,G\,M/c^2$). Especially promising in this sense are, in turn, binary radio pulsar systems containing a collapsed companion. In addition to the "direct" geodesics to the observer, the pulsar beam can be deflected by large angles by passing in the vicinity of the collapsed star, therefore reaching the observer along an "indirect" geodesics. This strongly deflected radio beam can be revealed through the characteristic orbital phase-dependent pulse arrival times or, equivalently, Doppler shifts (see also Parodi 1988).

In this paper we present a semi-classical derivation that allows to calculate approximate time delays and frequency shifts for the strongly deflected beam.

Our paper is organised as follows: in Section 2 we describe the trajectory and intensity reductions of strongly deflected light beams in a Schwarzschild metric and justify the approximations used in our semi-classical approach. Based on this, the orbital dependence of the pulse time delays and frequency shifts expected by an observer at infinity are worked out for the "indirect" path (Section 3). The application to the relativistic binary pulsars PSR 1913+16 and PSR 1534+12 is presented in Section 4. Our results are discussed in Section 5.

* e-mail: campana@astmim.mi.astro.it
† e-mail: stella@astmim.mi.astro.it

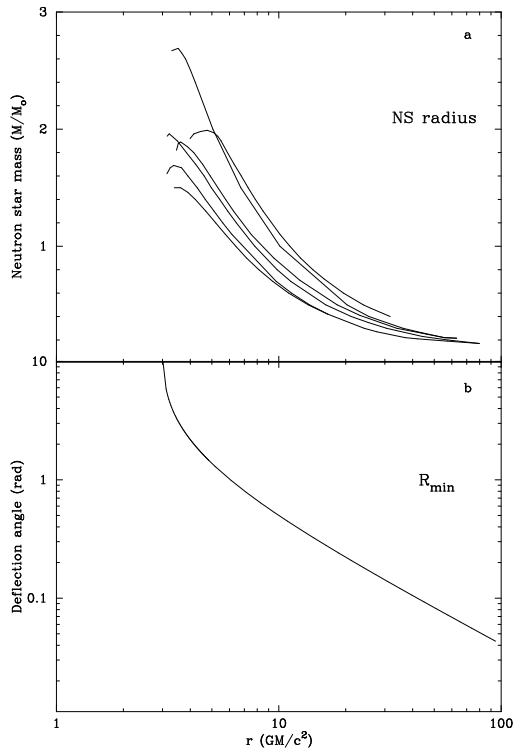

**Figure 1.** Mass-radius relation of neutron stars for different equation of state taken from Arnett & Bowers (1977) (panel *a*) and the total deflection angle $\theta$ as a function of the closest approach distance $R_{\min}$ (panel *b*). Note that radii are in units of $GM/c^2$.

## 2  STRONG FIELD LIGHT DEFLECTION IN THE SCHWARZSCHILD GEOMETRY

Collapsed stars, i.e. neutron stars and black holes, are the only known astrophysical objects that can gravitationally deflect light by large angles ( $\gtrsim 1$ rad). According to general relativity, the light deflection caused by a spherical, non-rotating star, can be calculated by integrating the null-geodesics equation for a Schwarzschild geometry

$$\frac{d^2 u}{d\varphi^2} = 3u^2 - u \,, \qquad (1)$$

where $u = GM/c^2 r$, with $r$ the radial coordinate, and $\varphi$ is the azimuthal angle in the plane of the photon trajectory (see e.g. Misner, Thorne & Wheeler 1973).

In panel *b* of Fig. 1 the deflection angle $\theta$ of a photon emitted from a large distance is plotted as a function of the radius of closest approach $R_{\min}$. Deflection angles in excess of $\sim 1$ rad are possible only if $R_{\min} \lesssim 6\,GM/c^2 \sim 9 \times 10^5 (M/M_\odot)$ cm. Besides black holes, only neutron stars, among known astrophysical objects, can be compact enough to satisfy the condition above.

In binary systems containing a collapsed star, photons emitted by the companion can in principle reach the observer following "indirect" geodesics characterised by large deflection angles. These geodesics are such that most of the deflection takes place in the strong field region of the collapsed star ($\sim 10\,GM/c^2$; see Fig. 2). If we limit ourselves to geodesics characterised by $\theta < \pi$, we are justified in adopting an approximation in which the "indirect" photons: *a)* first travel towards the compact object in a straight trajectory, *b)* undergo an instantaneous deflection at the compact object and *c)* proceed towards the observer along another straight trajectory (see dashed lines in Fig. 2).

In this semi-classical approach the deflection angle $\theta$ of the "indirect" photons is

$$\theta = \arccos\left(\sin i\,\sin(\phi + \omega)\right)\,, \qquad (2)$$

where $\phi$ is the position angle of the emitting star measured from periastron, $\omega$ the longitude of periastron and $i$ the system inclination.

A bundle of "indirect" photon geodesics spreads over a large solid angle after deflection, so that its intensity, $I$, is related to the intensity before deflection $I_0$ by the relation:

$$\frac{I}{I_0} = \frac{d\Omega_0}{d\Omega} = \frac{1}{1 - d\cos\chi\,\frac{d\theta}{db}}\,\frac{\sin\chi}{\sin\theta}\,, \qquad (3)$$

where $d\Omega_0$ and $d\Omega$ represent the solid angle subtended by the bundle of photon geodesics before and after deflection, respectively. Here $\chi$ is the angle between the direction of emission of the photon and the line connecting the emitting point to the compact object and $d$ the distance between the two stars; $b = d\sin\chi$ is the impact parameter.

To derive $I/I_0$ for given values of $d$ and $\theta$, $d\theta/db$ and $\chi$ must be calculated numerically by integrating Eq. 1. It is found that the intensity of the deflected signal is strongly modulated with the orbital phase, especially in the case of high inclinations and/or eccentricities. An order of magnitude estimate of the intensity reduction of the "indirect" signal is provided by $I/I_0 \sim (3\,GM/c^2\,d)^2$. For typical separations in compact binaries this gives values between $10^{-9}$ and $10^{-11}$.

Solutions to Eq. 1 can also be found which correspond to geodesics undergoing even larger deflection angles ($\theta' > \pi$). In this case the photons "revolve" $n$–times around the compact star close to $\sim 3\,GM/c^2$ before reaching the observer ($n = 1, 2, ..., \infty$). For these trajectories $\theta' \simeq 2\pi n - \theta$ or $\theta' \simeq 2\pi n + \theta$, depending on whether the photons revolve counterclockwise or clockwise in the plane of Fig. 2. These geodisics contribute to the observer an intensity which is many orders of magnitude lower than that of the "indirect" geodesics with $\theta < \pi$ (in total a factor of $< 10^{-3}$; Luminet 1979); therefore we neglect them in the rest of this paper.

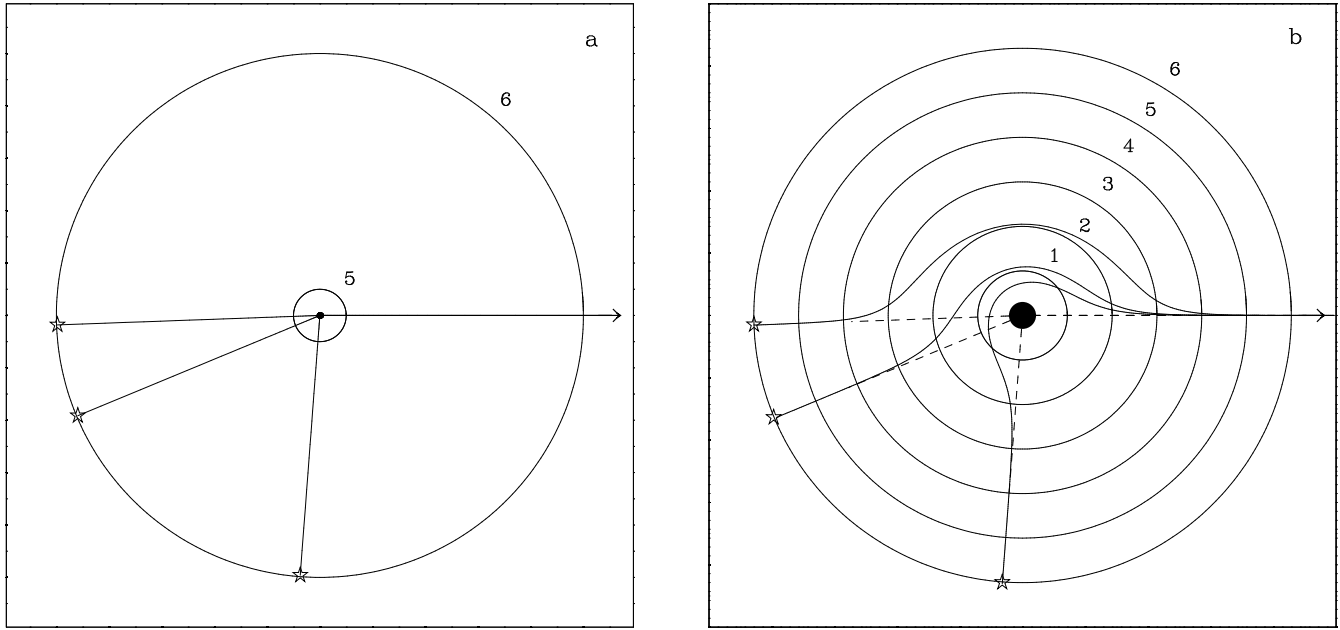

**Figure 2.** Three examples of photon geodesics in a Schwarzschild geometry. The photons emitted by star 1 at a distance of $10^6\, G\,M/c^2$ (star symbols) are deflected in the direction of the observer at infinity by the collapsed object at the center of the figure (star 2). In a linear scale representation (panel *a*) it is apparent that each photon trajectory is very well approximated by two straight segments, joining at the collapsed object. A logarithmic radial scale is used in panel *b* to emphasise the difference at small radii between the photon geodesics (solid lines) and the straight line approximation (dashed lines, see text). The collapsed object is represented here with a filled circle of radius of $2\,G\,M/c^2$. In both panels the circles centered on the collapsed object are marked with the logarithm of their radius in units of $G\,M/c^2$.

## 3 TIME DELAYS AND FREQUENCY SHIFTS OF STRONGLY DEFLECTED PHOTONS

Photons travelling along the "indirect" path are characterized by different time delays or, equivalently, Doppler shifts compared to "direct" photons. Based on our semi-classical approximation, the time delay $\Delta t_{tot}$ of an "indirect" photon is given by the sum of the light travel time $\Delta t_{12}$ from the emitting star (star 1) to the compact object (star 2) and the light travel time $\Delta t_{2\infty}$ from the compact object to the observer. In the classical limit these delays are:

$$\Delta t_{12} = \frac{d}{c} = \frac{a\,(1-e^2)}{c\,(1+e\cos\phi)}, \qquad (4)$$

$$\Delta t_{2\infty} = -\frac{a_2\,\sin i}{c}\,(1-e^2)\,\frac{\sin(\phi+\omega+\pi)}{1+e\cos\phi}, \qquad (5)$$

where $e$ is the eccentricity, $a_1$ and $a_2$ the semi-major axis of the orbit of the two stars, with $a = a_1 + a_2$. In the terminology used for delays in relativistic binary pulsars (e.g. Damour & Taylor 1992), this approximation accounts only for the Roemer and the Shapiro delays for the "indirect" beam. Compared to these the Einstein and aberration delays are expected to provide only minor corrections.

Similarly, the total Doppler shift of the "indirect" photons $z_{tot}$ can be calculated as:

$$1 + z_{tot} = (1 + z_{12})(1 + z_{2\infty}) \simeq 1 + z_{12} + z_{2\infty}. \qquad (6)$$

The first shift $z_{12}$ corresponds to the path from star 1 to star 2 and depends on the relative radial velocity of the two $\Delta v_{12}$:

$$z_{12} = \frac{\Delta v_{12}}{c} = \left(\frac{2\pi}{P_{orb}}\right)\frac{a}{c\sqrt{1-e^2}}\,e\,\sin\phi, \qquad (7)$$

where $P_{orb}$ is the orbital period. The second shift $z_{2\infty}$ derives from radial motion of the deflecting star relative to the observer:

Table 1. Approximate parameters of PSR 1913+16 and PSR 1534+12 used for Figures 3–6.

|  | PSR 1913+16 | PSR 1534+12 |
|---|---|---|
| Pulsar period $P$ (ms) | 59.030 | 37.904 |
| Distance (kpc) | 7.13[†] | 0.68[†] |
| Flux at 400 MHz (mJy) | 4 | 36 |
| Orbital period $P_{orb}$ (hr) | 7.752 | 10.098 |
| Projected semi-major axis $a_1 \sin i$ (lt s$^{-1}$) | 2.342 | 3.729 |
| Eccentricity | 0.617 | 0.274 |
| Time of periastron $T_0$ (MJD) | 46443.996 | 48262.843 |
| Longitude of periastron $\omega$ (deg) | 226.575 | 264.972 |
| $i$ (deg) | 47.223 | 80.401 |
| Pulsar mass $M_1$ ($M_\odot$) | 1.441 | 1.32 |
| Companion mass $M_2$ ($M_\odot$) | 1.387 | 1.36 |

[†] from Taylor & Cordes (1993).

$$z_{2\infty} = -\left(\frac{2\pi}{P_{orb}}\right) \frac{a_2 \sin i}{c\sqrt{1-e^2}} \left(e \cos\omega + \cos(\phi+\omega)\right) . \quad (8)$$

## 4  RELATIVISTIC RADIO PULSAR BINARIES: THE CASES OF PSR 1913+16 AND PSR 1534+12

In view of the very low intensity along the "indirect" path, the detection of strongly deflected photons emitted from an ordinary star in a single collapsed object binary appears to be highly unlikely. (Indeed, it would probably be difficult to reveal a spectral line feature dimmed by 3–4 orders of magnitude). In the case of a radio pulsar orbiting another collapsed object, the possibility of maintaining the pulse phase coherence over very long time intervals (tens of years) might drastically increase the sensitivity to the strongly deflected light beam. Therefore, these systems appear to be the most promising candidates for the detection of "indirect" signals.

The analysis of pulse arrival times from radio pulsars orbiting a companion star allows to measure the five Keplerian parameters (orbital period, projected semi-major axis, eccentricity, time of passage and longitude of periastron), like in single-component spectroscopic binaries. In general seven orbital parameters are needed to fully specify the dynamics of a binary system. Radio pulsar systems in which relativistic effects are important, notably double collapsed binaries, allow in principle to measure up to seven post-Keplerian parameters. Any two of these are sufficient to infer the star masses and system inclination, therefore determining completely the binary. Each of the remaining post-Keplerian parameters can be used to obtain a distinct test of general relativity (e.g. Taylor 1992). Thus far this has been possible for PSR 1913+16 and PSR 1534+12, two radio pulsar-neutron star binaries which provide very important and increasingly accurate tests of general relativity (for a review see Taylor et al. 1992). These binaries are also the best available candidates for the observation of the strong field light deflection discussed in this paper. Their approximate parameters are summarised in Table 1 (some of them were used to calculate the curves in Figs. 3–6).

One necessary condition for photons to be deflected at large angles by the compact object is that the radio pulsar beam sweeps sufficiently close to the companion star, at least during part of the orbit. For the purposes of our discussion it is sufficient to approximate the radio pulsar beam with a small cone, of semi-amplitude $\beta$, aligned with the magnetic field axis at a neutron star colatitude $\alpha$ (for more details see Lyne & Manchester 1988; Rankin 1993).

In the case of PSR 1913+16 the likely absence of geodetic precession suggests that the radio pulsar orbital and spin angular momentum vectors of are nearly parallel (Cordes, Wasserman & Blaskiewicz 1990). In this case the condition for the pulsar beam to sweep the companion star is $90° - \alpha \lesssim \beta$. On the other hand, it must be $|i - \alpha| \lesssim \beta$ because the "direct" radio pulsar beam reaches the earth. Pulse shape and polarisation measurements are used to derive approximate values for $\alpha$ and $\beta$. For PSR 1913+16 Cordes et al. (1990) and Rankin (1993) obtain $\alpha \sim 45°$ and $\beta \sim 17°$. Therefore it is unlikely that the radio pulsar beam of PSR 1913+16 intercepts the companion neutron star.

The geometry of PSR 1534+12 is much more favorable, because: *a)* the presence of an interpulse clearly indicates that the radio pulsar is a nearly "orthogonal rotator", with the line of sight close to the neutron star's equator; *b)* the binary system is characterised by a high inclination ($i \simeq 80°$). Indirect arguments suggest that the spin and orbital axes might be misaligned by a relatively small angle ($\sim 17°$; Wolszczan 1991). Even if no values of $\alpha$ and $\beta$ have yet been published, the observed large width and complex profile of the pulses testify to the presence of a relatively large beam, which is likely to sweep across the companion star. In particular, for PSR 1534+12 the condition that the radio pulsar beam sweeps the companion star at all orbital phases is $|(90° \pm 17°) - \alpha| \lesssim \beta$.

For the "indirect" beam to be deflected in the direction of the earth, the collapsed companion must be compact enough that its surface does not intercept the relevant pho-ton geodesics. Fig. 3 shows the deflection angle $\theta$ required for "indirect" photons to reach the earth at different orbital phases. In the case of PSR 1534+12 the deflection angles range from $\sim 3$ rad near inferior conjunction, to values as low as 0.2 rad close to superior conjunction, due to the high system inclination. A smaller range of deflection angles is required in the case of PSR 1913+16. It is apparent from Fig. 1 that, while deflection angles of up to $\sim 1$ rad are possible for most equations of state, a value of $\theta \sim 3$ rad is not achieved by a $\sim 1.4\,M_\odot$ neutron star.

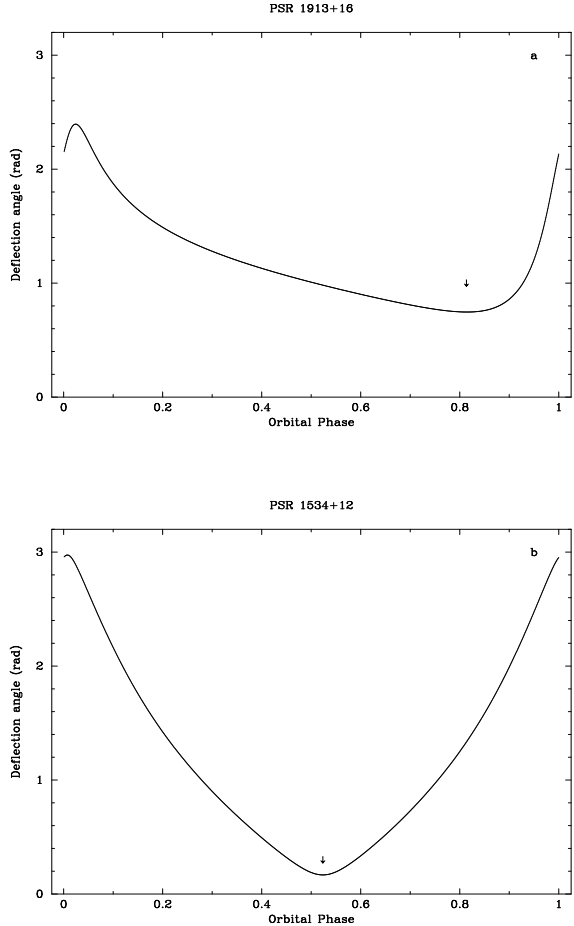

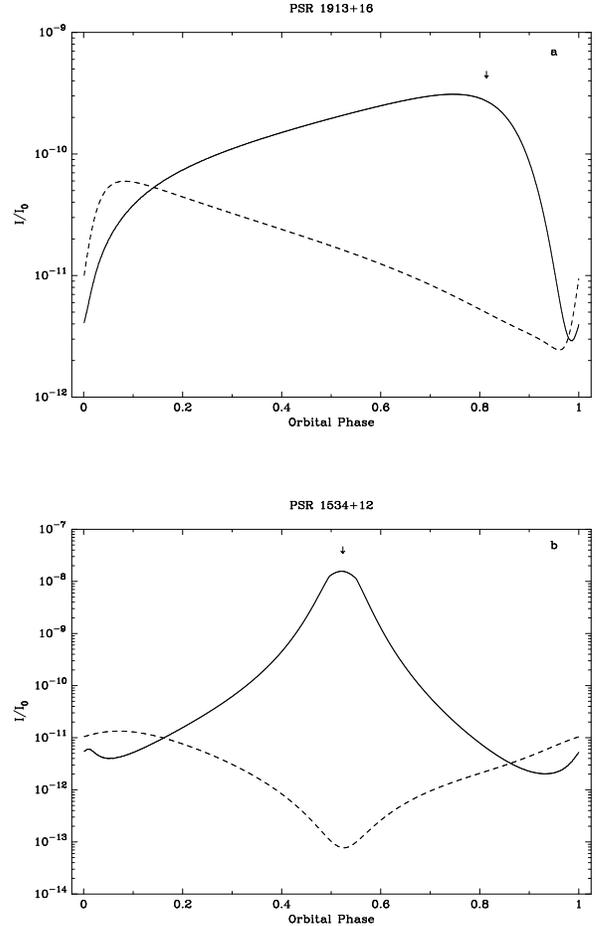

**Figure 3.** Deflection angle $\theta$ as a function of the orbital phase for PSR 1913+16 (panel *a*) and PSR 1534+12 (panel *b*). Here and in the following Figures the orbital phase is defined as $(t - T_0)/P_{orb}$ (see also Table 1) and the arrows indicate the phase of superior conjunction of the radio pulsar.

**Figure 4.** Intensity reduction of the "indirect" beam as a function of the orbital phase (solid lines): panel *a* is for PSR 1913+16, panel *b* for PSR 1534+12. The dashed lines give the intensity of photons reflected from the surface of the companion neutron star (see text).

Fig. 4 shows the reduction of the intensity of the strongly deflected beam as a function of orbital phase (see Section 2). PSR 1913+16 is characterized by a reduction factor $\lesssim 10^{-10}$ (panel *a*); on the contrary for PSR 1534+12 the reduction of the intensity of the deflected beam is limited to $2 \times 10^{-8}$ close to superior conjunction (panel *b*). The intensity of the strongly deflected beam relative to the "direct" beam can in principle be much higher than the values above if the pulsar beam geometry is such that emission in the direction of the collapsed companion is favored.

A possible problem derives from the the fact that, besides the gravitationally deflected beam, part of the photons reflected by the surface of the neutron star companion are also directed towards the earth. The intensity of the reflected beam can be roughly estimated as:

$$\frac{I_{refl}}{I_0} \sim \left(\frac{R_{NS}}{2\,d}\right)^2 (1 - \cos\theta) \qquad (9)$$

with $R_{NS} = 10^6$ cm the neutron star radius. The last term in parentheses corresponds to the fraction of the illuminated side of the neutron star which is visible from the earth. Eq. 9 neglects photon deflection and assumes an albedo of 1. The orbital dependence of the reflected intensity is shown by the dashed lines in Fig. 4. It is apparent that while the deflected and reflected beams can have intensities in the same range, their orbital dependence is drastically different. For PSR

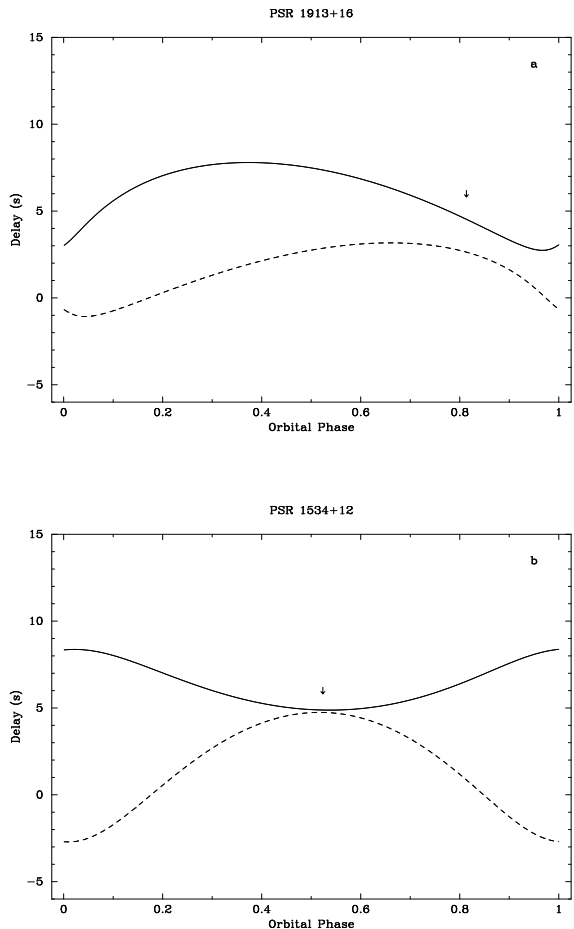
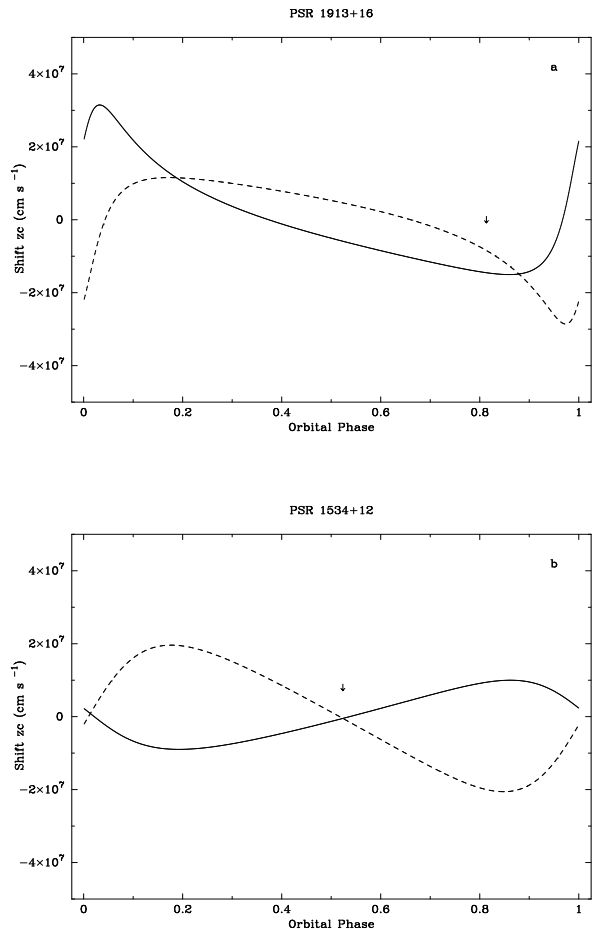

**Figure 5.** Time delays as a function of the orbital phase: the solid lines give the total delay of the "indirect" beam ($\Delta t_{tot}$); the dashed lines the delay of the "direct" beam. Panel *a* and *b* show the cases of PSR 1913+16 and PSR 1534+12, respectively.

**Figure 6.** Frequency shifts of the "indirect" (solid lines) and "direct" (dashed lines) beams as a function of the orbital phase. Panel *a* refers to PSR 1913+16; panel *b* to PSR 1534+12.

1534+12, in particular, due to the high inclination, the maximum intensity of the gravitationally deflected beam and the minimum intensity of the reflected beam occur close to the superior conjunction and are separated by more than 5 orders of magnitude.

In order to calculate the delays and the shifts of the "indirect" beam in the case of radio pulsar binary systems one further classical effect of order $P/P_{orb}$ has to be taken into account. This derives from the fact that the time interval between two consecutive sweeps of the pulsar beam across the companion star differs slightly from the neutron star spin period, due the angular displacement caused by the orbital motion. In the limit in which the orbital and spin angular momentum vectors are (nearly) coaligned, the total delay of the strongly deflected beam $\Delta t_{tot}$ contains an additional "synodic" correction, measured from the superior conjunction, of

$$t_{syn} = P\,\frac{\phi - \omega + \pi/2}{2\,\pi}\ . \tag{10}$$

Similarly, a term of

$$z_{syn} = \frac{P}{P_{orb}}\,\frac{\phi - \omega + \pi/2}{2\,\pi}\ , \tag{11}$$

should be added to the total shift on the right-hand side of Eq.6.

Figs. 5 and 6 show, respectively, the time delays and the frequency shifts of the "indirect" beam in the case of PSR 1913+16 (panels *a*) and PSR 1534+12 (panels *b*) as a function of the orbital phase. It is apparent that the time delays of the strongly deflected beam are somewhat larger than the delays of the "direct" beam, due to the presence of the term $\Delta t_{12}$. The frequency shifts of the "indirect" beam are of comparable amplitude and generally in antiphase with those of the "direct" beam. Note that close to the superior conjunction of PSR 1534+12, while the values of the delays and shifts of the deflected beam are similar to those of the "direct" beam, their orbital dependence is clearly different.

## 5 DISCUSSION

Based on a semi-classical approach, we have investigated the characteristics of strong gravitational field light deflection in binary systems containing a neutron star or a black hole. Photons originating from the companion star can be deflected by very large angles by passing in the vicinity of the collapsed object ($\gtrsim 3\,G\,M/c^2$) and therefore reach the earth. These "indirect" photons can in principle be identified through the pronounced and characteristic modulation of their frequency shifts or, equivalently, time delays throughout the orbit. The expected intensity of the strongly deflected beam, however, is very low and poses a very severe problem of detection.

In this context, binary systems containing a collapsed object and a radio pulsar represent the most likely candidates for revealing strong field light deflection. Indeed for some of these systems all relevant orbital parameters are accurately measured through the arrival time analysis of the ("direct") pulses and the application of the post-Keplerian formalism to relativistic binaries. This allows to make accurate predictions on the frequency shifts and time delays of the strongly deflected photons. The sensitivity to very weak signals can then be drastically improved by coherent summing of the pulse signal over very long time intervals.

We have discussed in some detail the cases of PSR 1913+16 and PSR 1534+12, currently the best studied radio pulsar relativistic binaries. Given the system geometry, the detection of the strongly deflected beam from PSR 1913+16 appears to be nearly impossible. Though still very difficult, the case of PSR 1534+12 is more promising. Owing to the high system inclination, a $\sim 0.2$ rad deflection is expected to take place around the superior conjunction of the radio pulsar; correspondingly the intensity of the "indirect" beam should be reduced by a factor of a few $\times 10^{-8}$, a much higher value than for other orbital phases. Moreover, if the radio pulsar emission pattern is concentrated in the direction of the companion star, the intensity of the deflected beam will be further enhanced.

The detection of the "indirect" beam would confirm for the first time the predictions of general relativity in the strong field regime ($\lesssim 10\,G\,M/c^2$), therefore opening a new perspective for the testing of alternative theories of gravity. In addition, an upper limit on the radius of the deflecting collapsed object can be obtained based on the distance of closest approach inferred from the strongly deflected light beam. In the case of a neutron star, this information can, in turn, be used to constrain the equation of state of high density matter.

*Acknowledgments*

We are grateful to A. Treves for pointing out an important omission at an early stage of this work, M. Colpi for her encouragement and Z. Arzoumanian (the referee) for useful comments. SC gratefully acknowledges support from an ASI fellowship. This work was partially supported through ASI grants.